\documentclass[aps,pra,reprint,groupedaddress,floatfix]{revtex4-1}
\usepackage{amsmath,amssymb}
\usepackage{mathtools}
\usepackage{graphicx}
\usepackage{physics}
\usepackage{slashed}
\usepackage{amsthm}
\usepackage{mathrsfs}

\usepackage[T1]{fontenc}
\usepackage{bm}
\usepackage{bbm}

\usepackage{dcolumn}

\usepackage{textcomp}
\usepackage[utf8]{inputenc}
\usepackage{tikz}
\usepackage{mathtools}

\usepackage{pgf}
\usepackage{verbatim}
\usepackage{import}

\usepackage{units}

\newcommand{\dop}{\text{DOP}}

\begin{document}

\title{Stopped and stationary light at the single-photon level inside a hollow-core fiber}

\author{Thorsten Peters}
\affiliation{Institut f\"{u}r Angewandte Physik, Technische Universit\"{a}t Darmstadt, Hochschulstrasse 6, 64289 Darmstadt, Germany}
\author{Ta-Pang Wang}
\affiliation{Institut f\"{u}r Angewandte Physik, Technische Universit\"{a}t Darmstadt, Hochschulstrasse 6, 64289 Darmstadt, Germany}
\author{Antje Neumann}
\affiliation{Institut f\"{u}r Angewandte Physik, Technische Universit\"{a}t Darmstadt, Hochschulstrasse 6, 64289 Darmstadt, Germany}
\author{Lachezar S. Simeonov}
\altaffiliation{Now at: Department of Physics, Saint Kliment Ohridski University of Sofia, 5 James Bourchier Boulevard, 1164 Sofia, Bulgaria}
\author{Thomas Halfmann}
\affiliation{Institut f\"{u}r Angewandte Physik, Technische Universit\"{a}t Darmstadt, Hochschulstrasse 6, 64289 Darmstadt, Germany}

\date{\today}

\begin{abstract}
An experimental platform operating at the level of individual quanta and providing strong light-matter coupling is a key requirement for quantum information processing. We report on narrowband light storage and retrieval as well as stationary light, based on electromagnetically induced transparency, for weak coherent light pulses down to the single-photon level with a signal-to-noise ratio of 59. The experiments were carried out with laser-cooled atoms loaded into a hollow-core photonic crystal fiber to provide strong light-matter coupling, thereby demonstrating the prospects for future quantum networks of such a platform.
\end{abstract}

\maketitle

\section{Introduction}
Optical quantum information processing \cite{NC01} requires generation, processing and detection of individual photons. Huge efforts have been dedicated to building high-brightness single-photon sources \cite{SL19}, implementing efficient quantum memories \cite{BGS15}, and developing schemes to provide strong light-matter coupling, allowing for interactions between individual photons \cite{CVL14}.

A powerful approach towards high memory
efficiency \cite{HTC18} and interaction between photons is based on
electromagnetically induced transparency (EIT) with atomic
ensembles \cite{FIM05}. 
EIT permits reversible group velocity control, e.g., for light storage and retrieval (LSR) \cite{FIM05} and stationary light pulses (SLPs) \cite{EHC19}, as well as interactions between light fields via a Kerr-type nonlinearity
\cite{LI01,AL02,ABZ05}. SLPs are of particular interest due to their potential for nonlinear optics at the few-photon level
\cite{ABZ05,CGM08,HCG12}. Although an EIT-driven 
memory and SLPs with non-classical quantum states were demonstrated in free-space cold atomic ensembles \cite{CMJ05,WLZ19,PCC18}, demonstrations of nonlinear optical interactions were so far restricted to
classical light pulses of large average photon number $\bar{n}$ \cite{CLH12}. Large nonlinearities require that two photons
interact with the same atom simultaneously, which can be quantified by the optical depth per atom $d_{opt}^*=\sigma_0/\pi w_0^2$ \cite{CVL14}, where $\sigma_0$ is the atomic absorption cross-section and $w_0$ is the waist of the Gaussian beam. This requires very tight focusing of the laser beams and thus results in small interaction volumes in free-space setups and,
hence, small optical depths $d_{opt}=\sigma_0 n L$, where $n$ is the
atomic number density and $L$ is roughly given by the Rayleigh
length. As the coupling strength of light to atomic ensembles is proportional to the optical depth $d_{opt}$, which also determines the LSR efficiency \cite{GAL07b}, experimental setups are required that provide both large $d_{opt}$ and $d_{opt}^*$. Therefore, in recent years there have been efforts to
couple atomic ensembles to waveguides such as hollow-core
photonic crystal fibers (HCPCFs) \cite{GBR06,TK07,BHB09}, tapered optical
nanofibers \cite{NSY18}, and nanoscale photonic
crystals \cite{GHY14}, which allow for tight transversal
confinement of light and atoms over macroscopic distances.

Operation at the quantum level requires strong
suppression of the background (caused mainly by the strong
control beam in EIT). However, spatial filtering as in free-space
setups cannot be applied due to the $1$D geometry of the
waveguides.
So far, LSR in waveguide-coupled atomic ensembles at the single-photon level (SPL) has been demonstrated for broadband photons employing a Raman protocol and room-temperature atoms \cite{SMC14} and narrowband photons employing EIT and cold atoms coupled to optical nanofibers \cite{GMD15,SCA15}.  
The latter demonstrations reported $d_{opt}\lesssim 6$ and a LSR efficiency of $\eta_{LSR}\lesssim 10~\%$ (where $\eta_{LSR}$ is defined as the ratio of retrieved to input pulse area).
Larger $\eta_{LSR}$ and the ability to generate SLPs is expected in HCPCFs filled with cold atoms due to the larger $d_{opt} \lesssim 1000$ in such setups \cite{BHP14}.
However, background suppression by polarization filtering is further
hampered in HCPCFs by strong birefringence \cite{BLK03,WLG05} stemming from structural features much smaller than the periodicity of the HCPCF cladding structure \cite{PBR05}.
Further difficulties arise also due to the correspondingly
larger control beam power $P_c\propto \sqrt{d_{opt}}$ required for
the same EIT transmission window width
$\Delta\omega_{EIT}=\Omega_c^2/\Gamma\sqrt{d_{opt}}$ \cite{FL02},
where $\Gamma$ is the excited state decay rate and $\Omega_c$ is
the control Rabi frequency.

Here we demonstrate the storage and retrieval of weak narrowband coherent light pulses, and the creation of SLPs down to the SPL in an ensemble of laser-cooled atoms inside a HCPCF. 
Compared to our previous work \cite{BSH16}, we carefully characterized the birefringence of our HCPCF to overcome its detrimental effects, enabling efficient polarization filtering. Also, we apply a fine structure transition with larger transition moment for the control, enabling operation at lower power, as well as reduced off-resonant absorption and inhomogeneous ac Stark shifts due to fewer excited hyperfine states at larger detuning.
This leads to an increased efficiency and improved signal-to-noise (SNR) ratio by roughly 20~dB and thus enables operation at the SPL.

\section{Experimental Setup}
\begin{figure}[t]
\centering
\includegraphics[width=\linewidth]{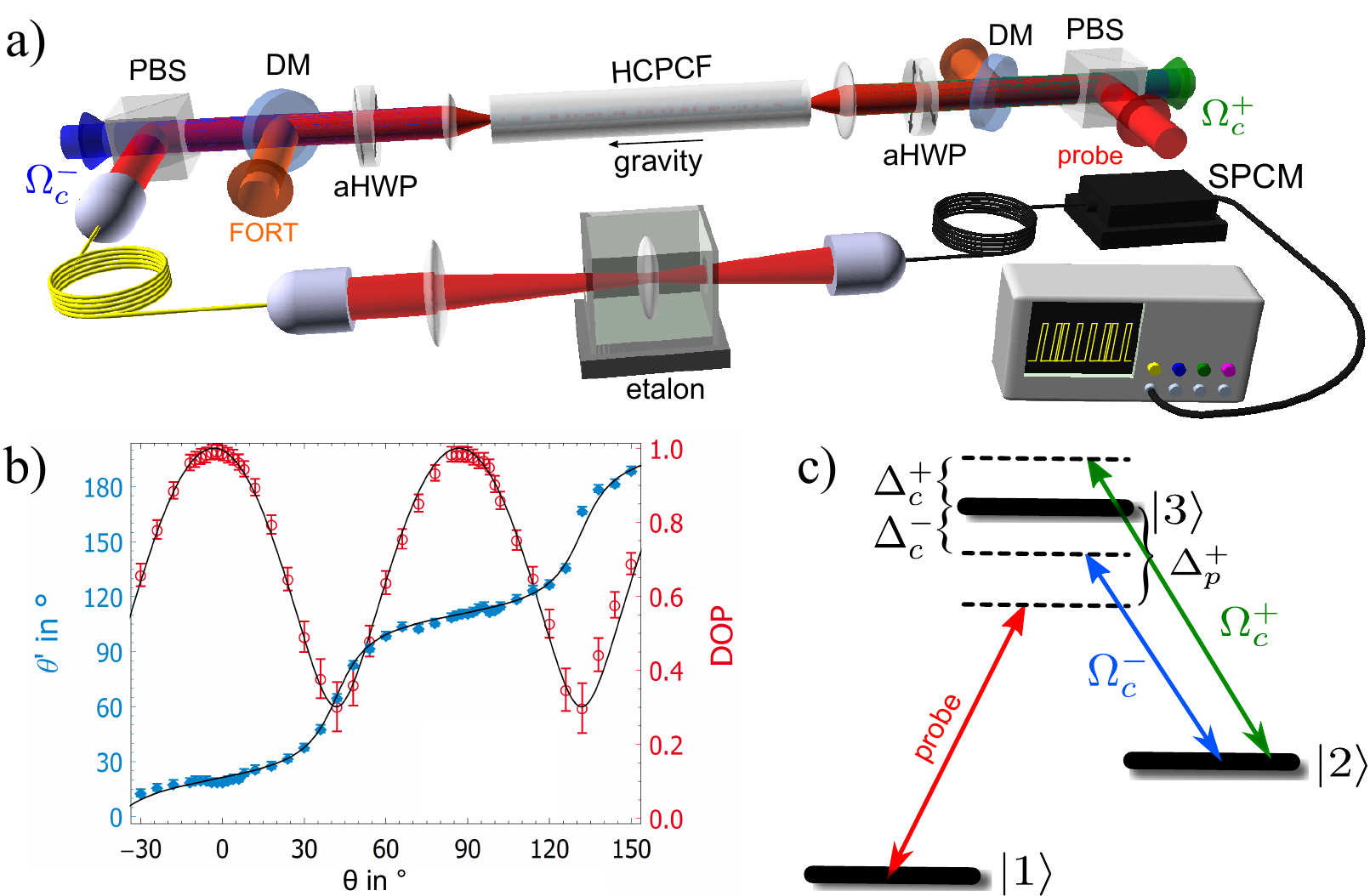}
\caption{\label{fig:setup} a) Simplified experimental setup. PBS: polarizing beam splitter, DM: dichroic mirror, SPCM: single-photon counting module. 
b) Measured DOP and output polarization direction $\theta'$ after the HCPCF (symbols) as a function of linear input polarization direction $\theta$. The black lines are simulations based on the theory in the SM. c) Level scheme for EIT-based LSR and SLPs. $|1,2\rangle= 5^2S_{1/2},\, F=1,2$, and $|3\rangle= 5^2P_{1/2},\, F'=1$.}
\end{figure}

We implement the experiment with cold $^{87}$Rb atoms, loaded from a magneto-optical trap (MOT) into a HCPCF (NKT Photonics HC-800-02), involving a far off-resonant optical trap (FORT) to guide the atoms inside the fiber and prevent collisions with the room-temperature fiber wall \cite{BHP14}. The probe laser is locked with an offset of 76~MHz to the corresponding transition via bichromatic saturation absorption spectroscopy \cite{GLH17}. The control laser is phase-locked to the probe yielding a two-photon linewidth of $\sim 8$ kHz \cite{BSH16}. Acousto-optic modulators (AOMs) shift the frequencies to the required detuning. Probe and control fields have orthogonal linear polarizations. 
The FORT laser system uses two superimposed orthogonally-polarized laser diode beams at 820~nm. With a trapping power of 110~mW inside the HCPCF, we obtain a trap depth of around 4~mK. 
 To adjust the linear polarizations of all fields inside the HCPCF at wavelengths of 780~nm, 795~nm, and 820~nm, we use achromatic half-wave plates (aHWPs) [see Fig.~\ref{fig:setup}(a)].
For more experimental details see \cite{BHP14,BSH16} and the Appendix.

\section{Experimental Results}
\textit{Suppression of the strong control beam}
Due to the HCPCF's strong birefringence \cite{BLK03,WLG05}, we previously could attenuate the strong control beam only by 35~dB using a combination of two PBSs behind the fiber. This lead to a total suppression of 74~dB (64~dB relative to the probe beam) \cite{BSH16} and limited the probe pulses to $\bar{n}\gtrsim 70$ photons. We therefore conducted a thorough study of the birefringence properties of our fiber by analyzing the degree of polarization (DOP) at the output of our HCPCF as a function of linear input polarization direction $\theta$.
Fitting the DOP and orientation $\theta'$ of the major polarization axis at the output [see Fig.~\ref{fig:setup}(b)] to theoretical predictions including linear and circular birefringence (see Appendix), we find that our fiber is predominantly linearly birefringent (probably due to the elliptical core) with only a small admixture of circular birefringence. The resulting polarization beat length is $L_b=8.0(8)~$mm, i.e., similar to standard polarization-maintaining fibers, with a maximum achievable DOP of $0.9925(10)$. Matching the input polarizations to the optical axis of the HCPCF
thus enables attenuation of the control beam by up to 86~dB, i.e., a  12~dB improvement, while using only a single PBS after the HCPCF. The overall transmission of the probe beam increases slightly to 0.13(1). By switching to the D$_1$ line with a larger transition strength for the control field, its background can be reduced by another 8~dB while maintaining a large Rabi frequency. This yields a total background suppression improvement of 20~dB and enables experiments at the SPL.

\begin{figure}[htbp]
\centering
\includegraphics[width=\linewidth]{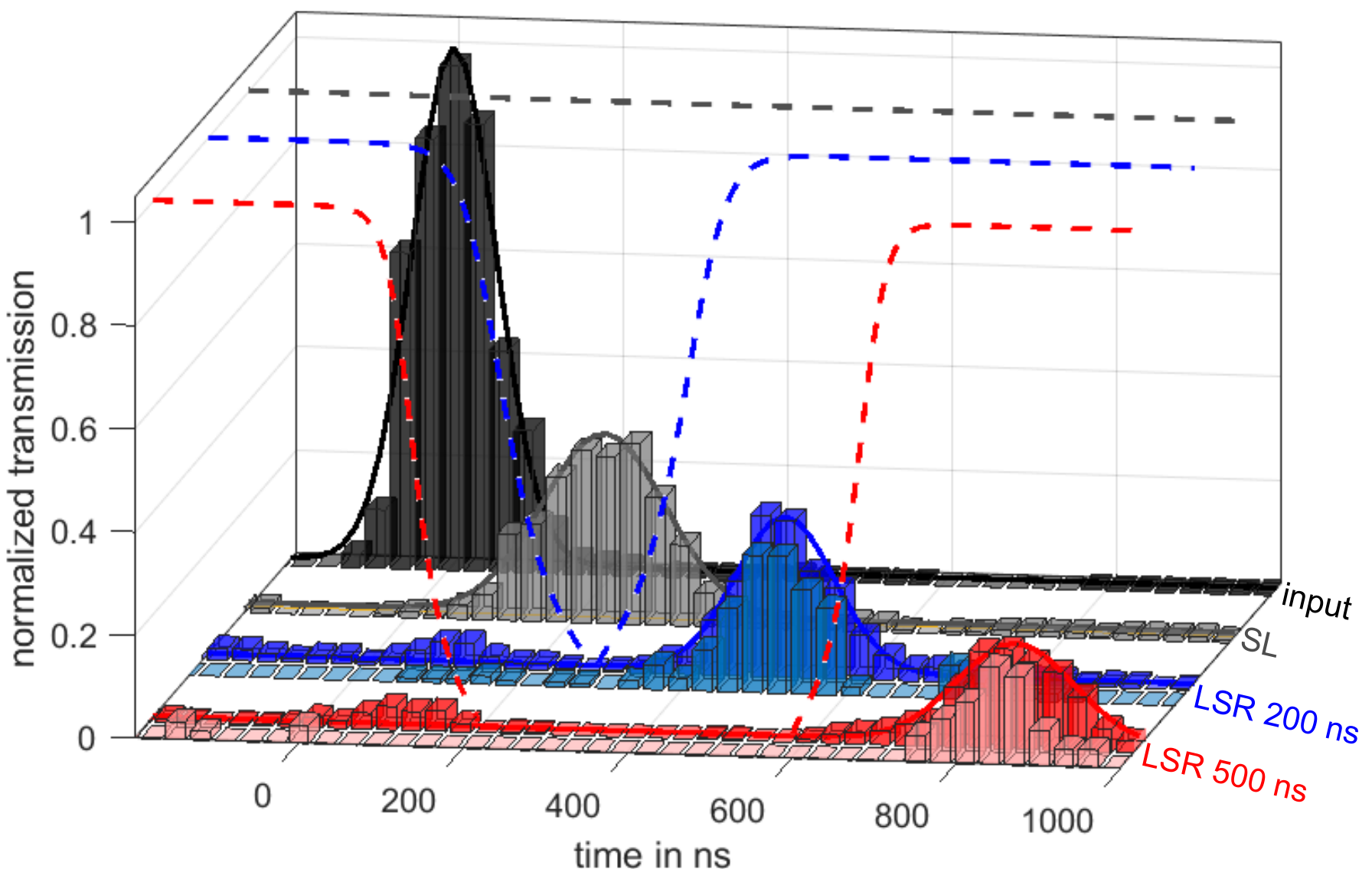}
\caption{\label{fig:LS}
Measured normalized transmission (bars) through the HCPCF vs. time for input pulses (black) containing $\bar{n}=17(3)$ photons, slow light (SL) (gray), and LSR for $\tau_{LSR}=200~$ns (blue) and $\tau_{LSR}=500~$ns (red). The light blue and red bars correspond to $\bar{n}=1.1(2)$ photons per input probe pulse. The solid colored lines are Gaussian least-squares fits to the corresponding experimental data. The dashed colored lines schematically show the corresponding timing of $\Omega_c^+(t)$. All experimental data are scaled with respect to the fit amplitude of the input pulse. The parameters are: $\Omega_c=3.6(2)\Gamma$, $d_{opt}=109(10)$. The data was averaged over 1500 runs.}
\end{figure}

\textit{Light Storage \& Retrieval of Coherent Light pulses} We now turn to LSR with the coupling scheme shown in Fig.~\ref{fig:setup}(c), where only the forward control field $\Omega_c^+$ is switched on, while the backward control field $\Omega_c^-\equiv 0$. Details on the LSR pulse sequence can be found in the Appendix. EIT creates dark-state polaritons (DSPs) with a moving photonic and a non-moving atomic coherence between states $|1\rangle$ and $|2\rangle$ \cite{FL00}.
The group velocity of the DSPs can be controlled by $\Omega_c^+$. Ramping $\Omega_c^+$ adiabatically down to zero maps the photonic DSP component onto the non-moving atomic component. By ramping the control field back up, the coherence is retrieved into a moving light pulse \cite{FL02}. We modulate the control field with an AOM to produce storage periods of variable duration $\tau_{LSR}$. 
The measured transmissions for $\tau_{LSR}=200$~ns and $\tau_{LSR}=500$~ns are shown in Fig.~\ref{fig:LS} in dark blue and red, respectively, for Gaussian probe pulses of $1/e$ full width $\tau_p=150(2)$~ns and $\bar{n}=17(3)$ photons per input pulse.
For reference we show the input probe pulse without atoms loaded into the HCPCF and a slow light pulse with constant control Rabi frequency in black and gray, respectively. The solid lines represent Gaussian least-squares fits to the experimental data. From these fits we obtain a LSR efficiency of $\eta_{LSR}=[0.36(4);\, 0.25(3)]$ for storage times of $\tau_{LSR}=(200; \, 500)$~ns, respectively. This is significantly larger compared to related work \cite{GMD15,SCA15}, due to the higher $d_{opt}$ in our experiment. By plotting the retrieval efficiency vs. storage time and fitting the data with an exponential decay we obtain a decay rate of $\gamma_{21}=0.26(2)\Gamma$.
In the same figure we also plot the results for LSR with coherent input pulses containing $\bar{n}=1.1(2)$ photons (light blue and red). There is very good agreement between the pulses at the SPL and those with higher photon numbers. 

\textit{Stationary Light Pulses.}
LSR allows for the stopping of light pulses. However, as the photons are converted into atomic coherences, no light is inside the medium during the storage period, prohibiting nonlinear optical interactions. In order to reduce the group velocity of the DSPs to zero while maintaining a photonic component (i.e., a light pulse), a second counter-propagating control beam of Rabi frequency $\Omega_c^-$ can be added while the probe pulse is inside the medium \cite{BZL03}. This creates a SLP with a quasi-stationary envelope which allows for nonlinear optical interactions. Details on the SLP sequence can be found in the Appendix. In order to keep the effective group velocity $v_{gr}\propto \Omega_0^2$ with $\Omega_0^2 = (\Omega_c^+)^2 + (\Omega_c^-)^2$ \cite{ZAL06} of the DSPs almost constant during the whole experiment, we reduce $\Omega_c^+$ during the SLP period $\tau_{SLP}$, when $\Omega_c^-$ is on. This allows for a better comparison with SL data, where $\Omega_c^- \equiv 0$. Without keeping $\Omega_0^2$ constant, the effective DSP group velocity would be more than doubled during $\tau_{SLP}$.
\begin{figure}[htbp]
\centering
\includegraphics[width=\linewidth]{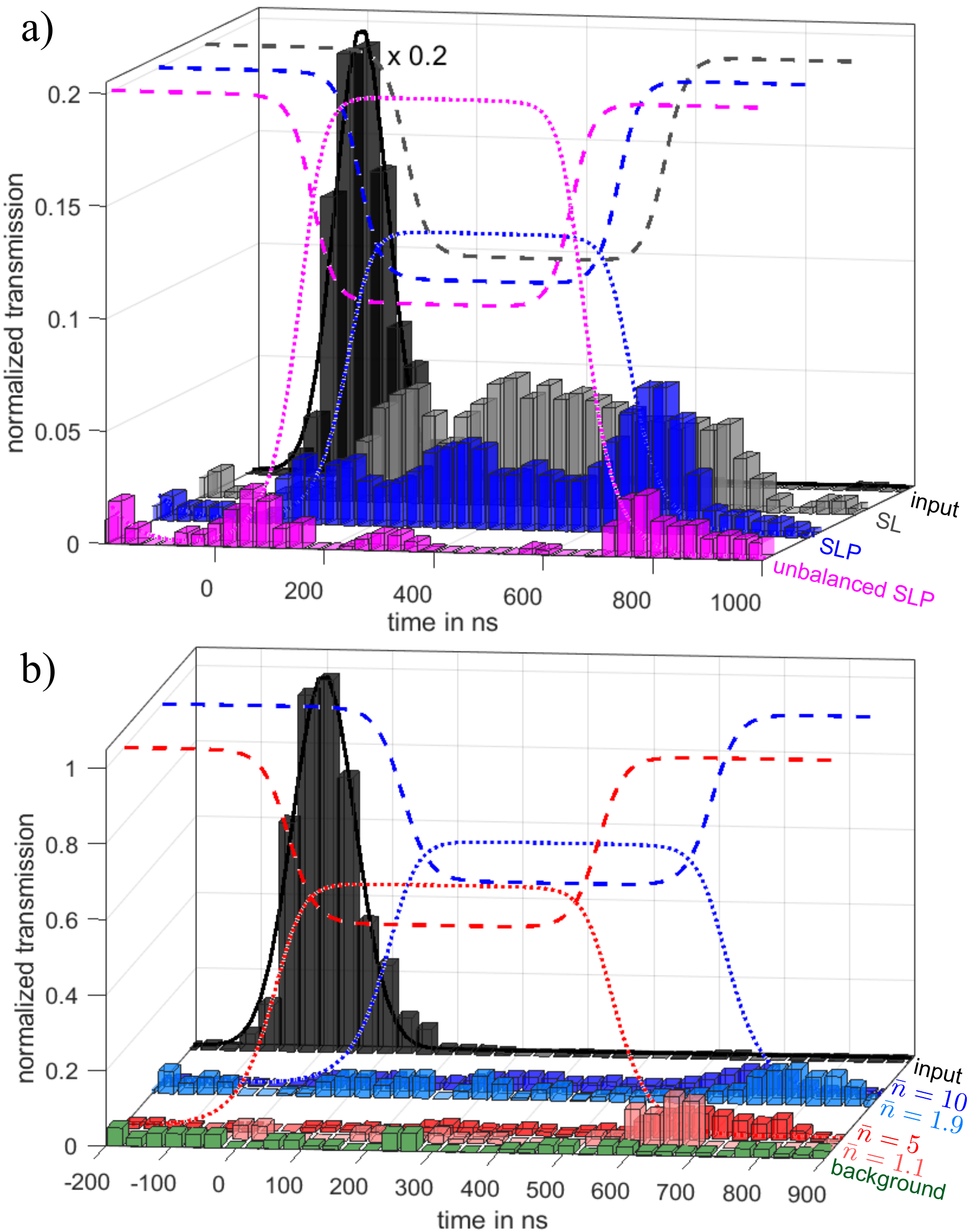}
\caption{\label{fig:SLP} 
Measured normalized transmission (bars) through the HCPCF vs. time. The solid black lines are least-squares fits to the Gaussian input pulse [black; scaled down by a factor of 0.2 in (a)] without atoms inside the HCPCF. All experimental data is scaled with respect to the fit amplitude of the input pulse. a) SL (gray) with a modulated $\Omega_c^+$ (gray dashed); SLP (blue) for $\Omega_c^+$ (blue dashed) and $\Omega_c^-$ (blue dotted) slightly unbalanced during $\tau_{SLP}$; backward propagating quasi-SLP (magenta) for $\Omega_c^+$ (magenta dashed) and $\Omega_c^-$ (magenta dotted) strongly unbalanced during $\tau_{SLP}$.
b) SLPs for $\bar{n}=10(2)$ [blue; same data as in (a)] and $\bar{n}=1.9(4)$ photons per pulse (light blue). Other experimental run for $\bar{n}=5(1)$ (red) and $\bar{n}=1.1(2)$ photons per pulse (light red) where the control timing was shifted by $-115~$ns compared to the blue data.
The background signal, when the input probe pulse is blocked, is shown in green.
The parameters are: $d_{opt}\sim 80(5)$ [except for the red data where $d_{opt}\sim 70(5)$], $\Omega_c^+=3.5(2)\Gamma$ (gray, blue, magenta dashed), $\Omega_c^-=[2.3(2); 3.5(2)]\Gamma$ [(blue; magenta) dotted], $\Omega_c^+=3.0(2)\Gamma$ (red dashed), $\Omega_c^+=2.4(2)\Gamma$ (red dotted) $\Delta_p^+=\Delta_c^+=0$, $\Delta_c^-=+2.5\Gamma$, $\tau_{SLP}=500$~ns.}
\end{figure}
In Fig.~\ref{fig:SLP}(a) we compare the results of three different conditions. The input pulse (black) is scaled down by a factor of 0.2 for better comparison. The probe signal for times $t\lesssim 150~$ns apparent in the gray, blue and magenta data sets corresponds to the rising edge of the probe pulses which have already left the medium before the SLP period starts. When $\Omega_c^- \equiv 0$ and $\Omega_c^+$ is modulated according to the gray dashed line, we observe a spread-out probe pulse due to the much reduced group velocity during $\Delta t_{SLP}=150~\textrm{ns}\lesssim t \lesssim 650~\textrm{ns}$ (gray bars). When the backward control field (magenta dotted line) is much stronger than the forward control (magenta dashed line) during  $\Delta t_{SLP}$, with $\Omega_c^-/\Omega_c^+=1.84$, we observe basically no transmission during $\Delta t_{SLP}$ and only a very small retrieved probe signal after the backward control is switched off (magenta bars). This small signal in addition is delayed with respect to the switch-off time of the backward control field. The delayed read-out can be well-explained by considering the SLP group velocity $v_{gr}^{SLP}=v_{gr} \cos 2\phi$ with $\tan^2 \phi= |\Omega_c^-|^2/|\Omega_c^+|^2$ \cite{ZAL06}. For $\Omega_c^-/\Omega_c^+=1.84$ the SLP propagates backward at around $-0.55 \times v_{gr}$ during  $\Delta t_{SLP}$ and thus exits the medium at later times during the retrieval in the forward direction. In contrast, when $\Omega_c^+$ and $\Omega_c^-$ are nearly balanced for optimum SLP conditions (blue dashed and dotted lines), there is some leakage of light during $\tau_{SLP}$ [but reduced compared to SL (gray bars)] and a significantly larger retrieved probe pulse area after the backward control is switched off (blue bars). This is the typical signature of a SLP having a non-vanishing intensity at the edges of the medium and therefore showing leakage \cite{BZL03,LLP09,BSH16,PCC18}. By calculating the retrieved pulse area after  $\Delta t_{SLP}$ we obtain a SLP retrieval efficiency of $\eta_{SLP}=0.14(2)$, i.e., a five-fold improvement compared to our previous work \cite{BSH16}. Therefore, $\eta_{SLP}$ is now of the same order as the corresponding $\eta_{LSR}$ (see above). The experimental data shown here are a selection of experimental runs where we varied the backward control Rabi frequency during $\tau_{SLP}$ between $0\Gamma\leq \Omega_c^- \leq 3.5\Gamma$. We obtained the largest SLP retrieval efficiency when $\Omega_c^- \sim 1.2\times \Omega_c^+$ during  $\Delta t_{SLP}$. We already discussed this in \cite{BSH16} as the result of the phase-mismatch present for the generated backward probe field in our 1D system. By choosing a larger  $\Omega_c^-$, this phase-mismatch can be compensated to some extend.
In Fig.~\ref{fig:SLP}(b) we compare the results for weak coherent input pulses containing $\bar{n}=10(2)$ (dark blue) to $\bar{n}=1.9(4)$ photons (light blue) and $\bar{n}=5(1)$ (dark red) to $\bar{n}=1.1(2)$ photons (light red) from a different experimental run. The background noise is shown for reference in green, reaching a level of $1.1\times 10^4$ photons per second and corresponding to 
a input SNR of roughly $59$. The efficiencies $\eta_{SLP}$ for higher and lower $\bar{n}$ agree with each other within $3\%$ for each experimental run. This data clearly demonstrates the potential of a HCPCF loaded with laser-cooled atoms as a suitable platform for EIT-based quantum information processing.

\textit{Discussion.} 
Although the demonstrated LSR and SLP retrieval efficiencies are sufficient to reliably detect pulses at the SPL, we might expect $\eta_{LSR}\gtrsim0.85$ for our $d_{opt}\sim 100$ \cite{GAL07b}.
To investigate this limited efficiency, we studied EIT spectra and compared the measured data to a numerical simulation including the radially varying control intensity and atomic density \cite{BSH16}.
Detrimental effects due to (inhomogeneous) ac Stark shifts are much reduced at the D$_1$ compared to the D$_2$ line. However, the measured peak transmission is lower than expected from the simulation.
For small EIT window widths, where typically LSR and SLP experiments are performed, this larger absorption simply reduces efficiencies. For larger window widths, however, this reduced transmission can be turned into a single low-transmission resonance within $\Delta\omega_{EIT}$, which we did not observe on the D$_2$ line for the same transition \cite{BSH16}.
This issue is currently still under investigation and, if solved, will lead to even higher storage efficiencies.

\section{Summary}
In summary, we demonstrated EIT-based LSR and SLPs for weak coherent light pulses containing
as low as $\bar{n}=1.1(2)$ photons per pulse, implemented in a medium of cold atoms inside a HCPCF. We observed a LSR and SLP efficiency of up to $0.36(4)$ and $0.14(2)$, respectively, at the SPL with a SNR of 59. 
This was enabled, amongst others, by a careful characterization and use of a mainly linearly birefringent HCPCF, resulting in efficient suppression of the strong control beam. With $d_{opt} \sim 100$ and $d_{opt}^* \sim 0.0017$, our results demonstrate the potential of HCPCFs loaded with laser-cooled atoms as a suitable experimental platform operating at the quantum level, while simultaneously providing strong light-matter coupling.

\section*{Funding Information}
European Union's Horizon 2020 research and innovation programme under the Marie Sklodowska-Curie grant agreement No. 765075.

\section*{Acknowledgments}
The authors thank H.R. Hamedi, G. Birkl and M. Fleischhauer for  discussions, F. Blatt for assistance with  measurements, and the group of T. Walther for providing us with a home-made ultra-low noise laser diode driver with high modulation bandwidth.

\appendix
\section{Birefringence characterization of the HCPCF}
\subsection{Theoretical analysis}
In analogy to common glass fibers, the HCPCF used in our experiment, is supposed to be highly birefringent due to the geometric asymmetry of the elliptic core and the additional symmetry breaking structural elements inside the cladding zone next to the core \cite{BLK03,WLG05}.
To model the birefringence of the fiber, we assume an in general elliptical birefringence. The initial light field is linearly polarized with real amplitude $\mathcal{E}_0$ and input orientation $\theta$, oscillating in the $x$-$y$-plane
\begin{equation}\label{eq:e0}
\vec{\mathcal{E}}(z=0) \equiv\vec{\mathcal{E}}_0 = \begin{pmatrix}\mathcal{E}_{0,x} \\ \mathcal{E}_{0, y} \end{pmatrix}
 = \begin{pmatrix}\mathcal{E}_0 \cos\theta\\ \mathcal{E}_0\sin\theta\end{pmatrix} .
\end{equation}
We use the Jones formalism with Jones matrix $J$ to describe the effect of the birefringent fiber on the electromagnetic field
\begin{equation}\label{eq:efield}
\vec{E}(z, t)=J \vec{E}(z=0,t),\qquad \vec{E}(z,t)=\textrm{Re} \left[\vec{\mathcal{E}}(z)e^{-i\omega t}\right],
\end{equation}
where we consider only the physically relevant real part.
The Jones matrix $J$ for elliptical birefringence reads \cite{Chen2005, Tabor1969}
\begin{equation}\label{eq:jonesmatrix}
J = \begin{pmatrix}
\cos\frac{\phi}{2}-i\sin\frac{\phi}{2} \cos\chi & -\sin\frac{\phi}{2}\sin\chi \\
\sin\frac{\phi}{2}\sin\chi & \cos\frac{\phi}{2}+i\sin\frac{\phi}{2} \cos\chi
\end{pmatrix},
\end{equation}
with $\phi$ defining the linear and $\chi$ the circular birefringence.

It is worth mentioning, that \eqref{eq:jonesmatrix} provides a fixed main axis. However, in the case of a HCPCF this theoretical main axis can differ from the geometrical main axis of the elliptical core [see Fig.~\ref{fig:fiber}(a)]. This can be possibly due to a change of its orientation over the fiber length, e.g., by an intrinsic torsion. We take this variation into account by adjusting the input angle $\tilde{\theta} =\theta-\beta$ with the orientation of the theoretical fixed main axis in relation to the horizontal axis $\beta$, visualized in Fig. \ref{fig:fiber} (a).

To characterize the birefringence of the HCPCF the orientation of the output field's main axis can be calculated with
\begin{equation}\label{eq:theta}
\theta' =\textrm{Re} \left[ \arctan \left( \frac{\mathcal{E}_y(\tilde\theta,\phi,\chi)}{\mathcal{E}_x(\tilde\theta,\phi,\chi)}\right)\right]+\beta
\end{equation}
as well as the degree of linear polarization (DOP). In general this is given by the relation of the field intensity's maximum and minimum
\begin{equation}\label{eq:dop0}
\dop = \frac{I_\text{max}-I_\text{min}}{I_\text{max}+I_\text{min}},
\end{equation}
with $\dop=1$ for completely linear polarized light.

We would like to express $\dop$ as a function of $\theta$. Therefore, we first calculate the general form of $\vec{E}(z,t)$. By inserting \eqref{eq:e0} and \eqref{eq:jonesmatrix} into \eqref{eq:efield} we get
\begin{align}
E_x(z, t) &= A \cos(-\omega t) + B \sin(-\omega t),\\
E_y(z, t) &= C \cos(-\omega t) + D \sin(-\omega t),
\end{align}
with
\begin{align}
A&= \mathcal{E}_0\left[\cos\tilde\theta\cos\tfrac{\phi}{2} -\sin\tilde\theta\sin\chi\sin\tfrac{\phi}{2}\right],\\
B&= -\mathcal{E}_0\cos\chi\sin\tfrac{\phi}{2}\cos \tilde\theta,\\
C&= \mathcal{E}_0\left[\cos\tilde\theta\sin\tfrac{\phi}{2}\sin\chi +\sin\tilde\theta\cos\tfrac{\phi}{2}\right],\\
D&= \mathcal{E}_0\sin\tilde\theta\cos\chi\sin\tfrac{\phi}{2}.
\end{align}
$\vec{E}(z,t)$ fulfills the equation of a rotated ellipse
\begin{equation}\label{eq:ellispe}
\frac{C^2+D^2}{\Delta^2}E_x^2 + \frac{A^2+B^2}{\Delta^2}E_y^2-2\frac{AC + BD}{\Delta^2}E_xE_y=1,
\end{equation}
with $\Delta=
AD-BC$.
To identify $I_\text{max}$ and $I_\text{min}$ we have to bring \eqref{eq:ellispe} to the normal form $r_1 E_x^2+ r_2 E_y^2 =1$, because then
\begin{equation}
\dop=\frac{r_2-r_1}{r_2+r1}= \sqrt{1-\frac{4(AD-BC)^2}{A^2+ B^2 +C^2+D^2}}.
\end{equation}
With $A^2+ B^2 +C^2+D^2= \mathcal{E}_0^2$ we finally obtain
\begin{equation}\label{eq:dop}
\dop\! =\! \sqrt{\!1\!-\! 4\cos^2\!\chi\sin^2\!\tfrac{\phi}{2}
\left(\sin\! 2\tilde\theta \cos\!\tfrac{\phi}{2} + \cos\! 2\tilde\theta \sin\!\tfrac{\phi}{2} \sin\!\chi \right)^2}.
\end{equation}

We can use this theoretical model for the DOP in \eqref{eq:dop} as well as the orientation of the output field in \eqref{eq:theta} to fit the experimental data and obtain the linear and circular birefringences.

\subsection{Experimental analysis}
Starting with an almost perfectly linear-polarized input field [$\dop = 0.999985(5)$], we measure $\theta'$ and $\dop$ depending on the input orientation $\theta$, adjusted by an aHWP, which impairs the input polarization to $\dop \geq 0.9984$.
The spatial intensity distribution of the output field is analyzed with a Glan-Thompson polarizer and normalized to the input power, resulting in $P(\vartheta)/P_0=b + a\sin(\vartheta)$. Then the DOP is given by $\dop=a/b$. The orientation $\theta'$ corresponds to the position of the maxima, respectively, the minima plus $90^\circ$, as the determination of the latter is more precise. Due to the high asymmetry of our fiber core it is important to make sure that only the approximately Gaussian-shaped mode transmitted through the core is analyzed. Higher-order modes that can propagate through the cladding zone drastically reduce the $\dop$.
\begin{figure}[t]
\centering
\fbox{\includegraphics[width=\linewidth]{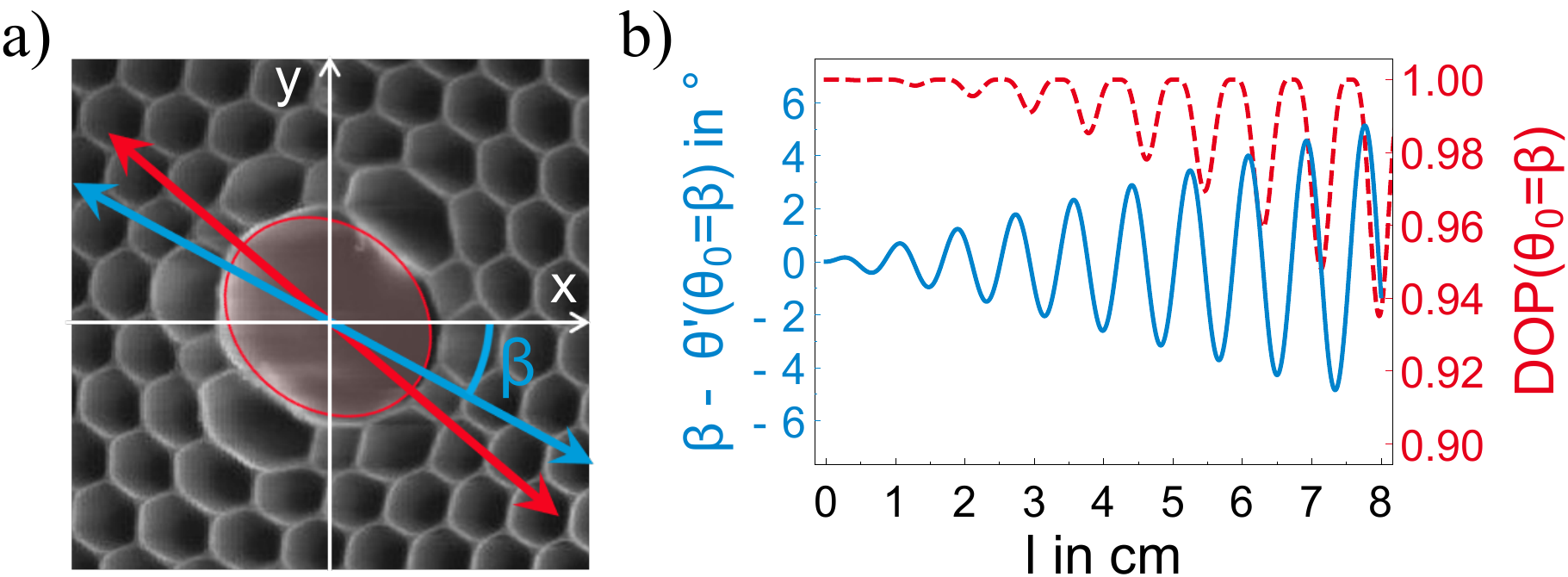}}
	\caption{(a) Scanning electron image of the central region of our HCPCF. The orientation of the optical axis $\beta$ (blue) can differ from the main axis of the elliptical core (red). (b) Calculated DOP as a function of fiber length $l$ and absolute rotation of the input polarization using the same parameters as in (b). For a length of $l\leq 3.5\,\text{cm}$ and an input angle $\theta=\beta$ the polarization can be maintained sufficiently well, if we assume $\dop \geq 0.99$ and $\Delta \theta\leq 2^\circ$.}
	\label{fig:fiber}
\end{figure}

The results for a nearly untwisted fiber (NKT Photonics HC-800-02) of length $l_\text{F}\approx22\,\text{cm}$ are depicted in Fig.~\ref{fig:setup}(b). Note that in general $\phi = \phi + 2n \pi,\ n \in \mathbb{Z}$. However with the wavelength-sweeping technique \cite{Kikuchi1983}, the exact value of $\phi$ can be calculated. Hence, we require two measurements of the $\dop$ for different wavelengths $\lambda_1\neq\lambda_2$ with $\dop(\theta,\lambda_1) \approx \dop(\theta,\lambda_2) $. More details can be found in \cite{Kikuchi1983}. Unfortunately, the latter condition is not fulfilled for a fiber length of $l_\text{F}\approx22\,\text{cm}$. We therefore used a shorter fiber of length $l_\text{F}\approx2\,\text{cm}$ to determine $\phi$. We then extrapolated $\phi(l_\text{F}\approx 22\,\text{cm}) \approx \frac{22\,\text{cm}}{2\,\text{cm}}\phi(l_\text{F}\approx2\,\text{cm})=11\cdot 15.20(6)=167(1)$. This value is then used as a starting point to fit the theoretical curves from \eqref{eq:dop} and \eqref{eq:theta} to the experimental data shown in Fig.~1(b) of the main document while slightly changing $\phi =167$ by $\pm \pi$. Here, the importance of the fit of $\theta'(\theta)$ becomes apparent. Only both fits together identify one unique $\phi$ in the given range. For our fiber we obtain the values
\begin{equation*}\label{eq:params}
\phi = 164.84(5),\quad \chi=0.50(5),\quad\beta= 0.15(2).
\end{equation*}

The degree of linear polarization obviously depends strongly on the input polarization orientation. The maximum $\dop_\text{max} = 0.9925(10)$ for an input orientation $\theta=-2^\circ \pm 1^\circ$ and $\theta'=19^\circ\pm1^\circ$ shows, that for certain input orientations, the HCPCF can almost maintain the input $\dop$. In addition, for an input orientation along the theoretical optical axis, we may expect to achieve the best conservation of the input orientation. However, for $\theta=\beta= 8.6^\circ\pm 1.1^\circ$ we get $\dop=0.93$ and $\theta' =24^\circ \pm 2^\circ$, because both depend on the fiber length. In Fig. \ref{fig:fiber} (b) the deviation between output and input polarization direction for $\theta =\beta$ and the $\dop$ is depicted as a function of the fiber length. For $l\lesssim 3.5\,\text{cm}$ indeed we can reach $\dop \geq 0.99$ and the change of the orientation $\Delta \theta \leq 2^\circ$ is very small. For fiber lengths $l\gtrsim 3.5~$cm a good DOP can be maintained by choosing a different input polarization orientation $\theta\neq\beta$, which is necessary due to the small elliptical birefringence of our fiber. Comparing the polarization beat length $L_b = 2\pi l_\text{F}/\phi\approx0.80(8)\,\text{cm}$, to common glass fibers with typical $L_b$ of several meters and polarization maintaining fibers with $L_b\lesssim 1 \dots 3\,\text{mm}$, in fact, our HCPCF can be assigned rather to the latter kind.

We note that the dominant linear birefringence of our fiber is probably due to the elliptical core with an aspect ratio of around 1.3 \cite{PBR05}. Without this large asymmetry, the birefringence properties might be more complex \cite{WLG05} and we would not be able to maintain the linear polarizations of the laser beams as they propagate through the HCPCF. This would inhibit efficient polarization filtering at the output and prevent experiments at the SPL.

\section{Experimental Sequences}
\begin{figure}[ht!]
\centering
\includegraphics[width=\linewidth]{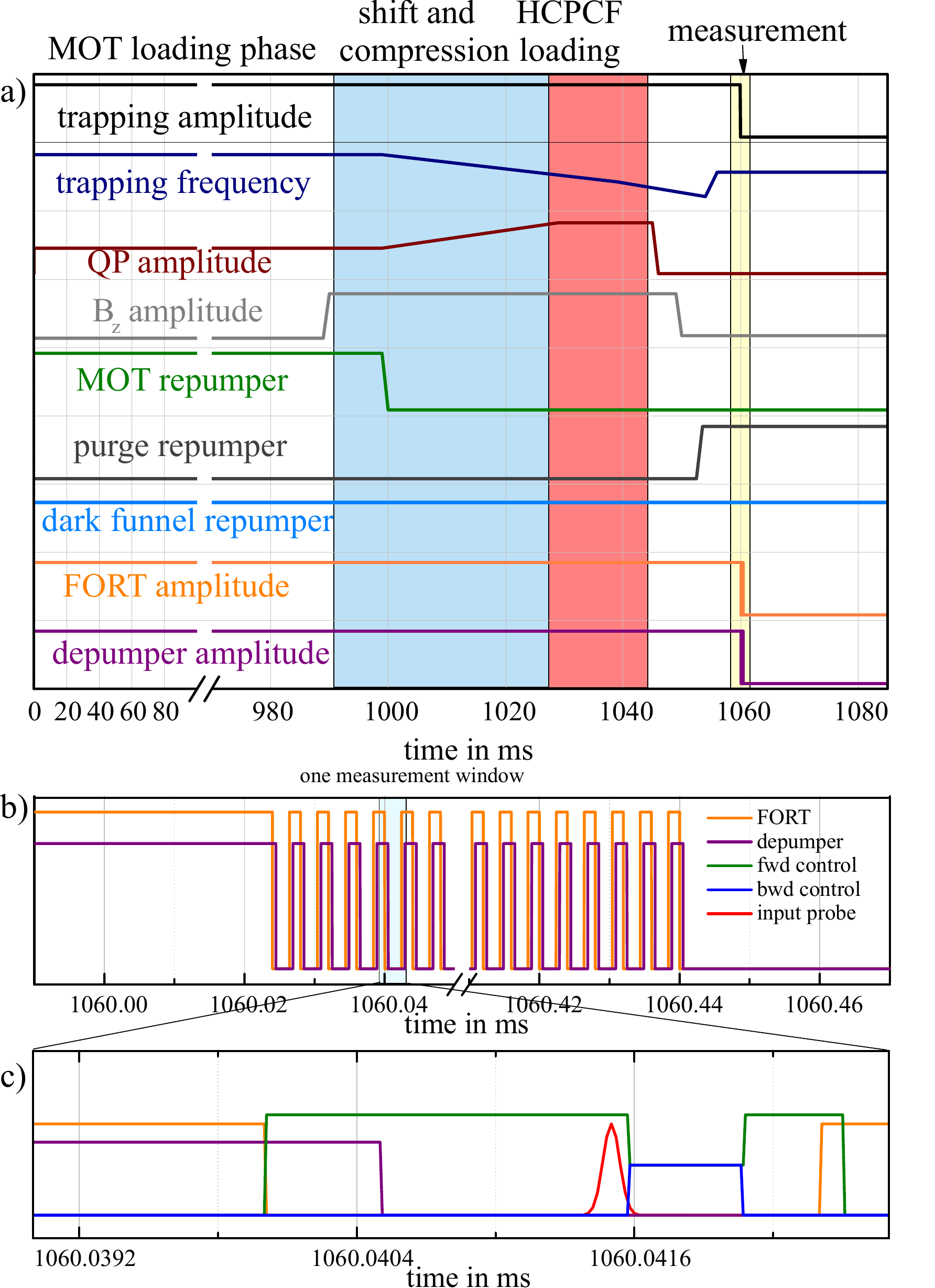}
\caption{\label{fig:sequence} a) Time sequences of laser and magnetic fields for loading the HCPCF with atoms. QP: quadrupole field of the MOT coils; $B_z$: magnetic offset field in the vertical direction that determines the zero point of the magnetic field and thereby shifts the atom cloud above the HCPCF. b) Sequence of the FORT and depumper for the SL/LSR/SLP experiments. c) Sequence of the laser fields used for SL/LSR/SLP measurements.}
\end{figure}

The schematic time sequence for loading the HCPCF with atoms is shown in Fig~\ref{fig:sequence}(a). The MOT is loaded for 990~ms with a Gaussian-shaped repumper beam in addition to the dark funnel repumper \cite{BHP14} to increase the number of atoms in the MOT. After the MOT loading phase the MOT repumper is switched off by a mechanical shutter and only the dark funnel repumper is left for increasing the density of atoms above the HCPCF. Simultaneously, the current of the quadrupole coils is doubled for compression of the cloud and the magnetic offset field in the $z-$direction (along the fiber axis) is changed to move the atoms towards the fiber tip. During the 40~ms long shift and compression and the 20~ms long HCPCF loading phases, the trapping detuning is ramped from $-2.5\Gamma$ to $-5\Gamma$ for sub-Doppler cooling above the HCPCF to increase the loading efficiency. During the HCPCF loading phase the atoms are held above the HCPCF where the FORT potential is strong enough to guide the atoms into the fiber. A depumper beam, tuned 133~MHz to the blue side of the transition $^2$S$_{1/2},\, F=2 \leftrightarrow ^2$P$_{3/2},\, F'=2$, to account for the ac Stark shift by the FORT, is continuously on and serves to confine the atoms inside the dark funnel in $F=1$ for enhancing the density near the fiber tip.
After loading the HCPCF is finished at 1050~ms and another 10~ms period for letting the quadrupole field decay, experimental runs of SL, LSR or SLPs start. The FORT (and the depumper) are rapidly switched on and off at a frequency of 250~kHz according to Fig.~\ref{fig:sequence}(b). This eliminates ac Stark shifts by the FORT during the measurement windows of 2.5~$\mu$s duration, but is still short enough to not lose a significant amount of atoms during this free expansion. Up to 50 measurements can be done with the same atoms loaded into the fiber, before losses became noticeable. In each time slot when the FORT is off, SL/LSR/SLP measurements [see Fig.\ref{fig:sequence}(c)] are performed. The first 4 time slots are used to prepare the population in $F=1$ by the control beam. We do not try to prepare the population in a single Zeeman level. 
Then, Gaussian probe pulses of $\tau_p=150(2)$~ns $1/e$ full width are sent into the HCPCF in the next 25 time slots. Afterwards, another 25 time slots are used to record the background signal without probe.
Each HCPCF loading cycle is followed by a reference measurement without atoms inside the HCPCF, by keeping the MOT quadrupole field off. We therefore can detect the background due to the laser fields and due to scattered photons from the atoms inside the HCPCF independently.

\section{LSR \& SLP Data Acquisition \& Analysis}
The transmitted photons are detected by a single-photon counting module (SPCM, PerkinElmer, SPCM-AQRH-12). The TTL output pulses of the SPCM are recorded with a digital oscilloscope (100 MHz bandwidth). The oscilloscope is triggered by the start of the first probe pulse. All events are then recorded during one measurement sequence lasting 200~$\mu$s (corresponding to 50 measurement windows separated by 4~$\mu$s) where the FORT is modulated for each single HCPCF loading cycle. In addition, the data is averaged over 60 HCPCF loading cycles.
The temporally resolved transmission signal is then generated via software analysis by slicing the data into bins of 30~ns width and averaging over the different measurement time slots as well as different HCPCF loading cycles. The same applies to the background signal which is subtracted from the recorded data.
The magnitude of the background signal for the measurements shown in Fig.~\ref{fig:SLP}(b) is $1.1\times 10^4\, (0.7\times 10^4)$ photons per second with (without) atoms inside the HCPCF.
%

\end{document}